\documentclass{article}
\usepackage{frascatiphys_R}

\def\pnn{$K \rightarrow\pi\nu\overline\nu$}

\def\kpnn{$K^+\rightarrow\pi^+\nu\overline\nu$}

\def\kpimue{$K^+ \rightarrow \pi^+ \mu^+ e^-$}

\def\kethree{$K^+ \rightarrow \pi^0 e^+ \nu$}
\def\kmuthree{$K^+ \rightarrow \pi^0 \mu^+ \nu$}

\def\klpnn{$K_L\rightarrow\pi^0\nu\overline\nu$}
\def\klpiee{$K_L \rightarrow \pi^0 e^+ e^-$}
\def\klpimm{$K_L \rightarrow \pi^0 \mu^+ \mu^-$}
\def\klpill{$K_L \rightarrow \pi^0 l^+ l^-$}
\def\klmue{$K_L \rightarrow \mu^{\pm} e^{\mp}$}
\def\klmumu{$K_L \rightarrow \mu^+ \mu^-$}

\def\kleegg{$K_L \rightarrow e^+ e^- \gamma \gamma$}
\def\klmmgg{$K_L \rightarrow \mu^+ \mu^- \gamma \gamma$}
\def\klpimue{$K_L \rightarrow \pi^0 \mu^{\pm} e^{\mp}$}
\def\kleemmlfv{$K_L \rightarrow e^{\pm} e^{\pm} \mu^{\mp} \mu^{\mp}$}
\def\klgg{$K_L \rightarrow \gamma \gamma$}

\def\ksppp{$K_S \rightarrow \pi^0 \pi^0 \pi^0$}
\def\kspiee{$K_S \rightarrow \pi^0 e^+ e^-$}
\def\kspimm{$K_S \rightarrow \pi^0 \mu^+ \mu^-$}
\def\kspill{$K_S \rightarrow \pi^0 l^+ l^-$}

\def\br{$\mathcal{B}$}

\begin{document}
\title{RARE KAON DECAYS: IL BUONO, IL BRUTTO, IL CATTIVO\\
}

\author{G. Redlinger\\
\em Brookhaven National Laboratory, Upton, NY, USA 11973
}
\maketitle
\baselineskip=11.6pt
\begin{abstract}
I briefly review recent progress in rare kaon decays,
where I take ``rare''
to mean those with $\mathcal{B} < \mathcal{O}(10^{-7})$.
\footnote{I set the scale by the
``classic'' example of rare, the
probability of getting killed by lightning in one year in the U.S.
See for example http://mathforum.org/dr.math/faq.}

\end{abstract}
\baselineskip=14pt
\section{Introduction}
The title of this talk (which by the way borrows from the original title of the
famous ``spaghetti Western'' ``The Good, The Bad, and The Ugly'') arose
from the convergence of several trains of thought: first, that this
conference was being held near Rome, where the filmmaker Sergio Leone
was born and lived; second, various allusions to current world
politics, which are best left to the coffee break; lastly,
and most importantly, I was inspired by a writeup of a talk by
Wilczek\cite{Wilczek}:

\begin{quote}
  ``Our current, working description of
  fundamental physics is based on three conceptual systems... it is
  not inappropriate to call them the Good, the Bad, and the Ugly''
\end{quote}

\noindent For the purposes of this talk, we concentrate on the ``Ugly'' which
is the flavor sector whose many parameters illustrate our lack of
understanding of the Higgs Yukawa couplings.
Experiment is a key driving force in making progress here; kaon decays have
had a glorious history in elucidating this physics, and continue to
serve as sensitive probes.

In the following, I cover rare kaon decay results\footnote{All
  quoted limits are at 90\% CL.} from approximately 2003 to the present.
  These can be grouped as follows:
\begin{itemize}
  \item The Good. This includes the study of signatures explicitly
    beyond the Standard Model (BSM), of which the best known are the
    lepton flavor violating decays.
    Here we also include modes sensitive to quark-mixing parameters
    and CP violation (CPV) with small theoretical uncertainty, thus
    making them 
    excellent candidates for searches for BSM physics.
  \item The Bad.  This include studies of the low-energy behavior of
  the strong interactions.  Obviously this is not ``bad'' in itself,
  but is not directly connected to studying the flavor sector.
  \item The Ugly.  These are decay modes that potentially probe
  quark-mixing, CPV and BSM physics but which do not lend
  themselves to a clean extraction of the fundamental parameters.
\end{itemize}

\noindent
I do not cover high-sensitivity experiments that require large numbers
of kaons, but where the underlying branching ratios are relatively
large, such as $\epsilon'/\epsilon$ or the search for T violation in
\kmuthree. Needless to say, in the limited space available I can only give a
cursory overview of the field; the interested reader should consult
the many reviews available.\cite{reviews}

\section{Lepton flavor violation}

The state-of-the-art in lepton flavor violation (LFV) is set by the BNL
experiments E871 and E865
which searched for the decays \klmue~ and \kpimue~ respectively. The
two decay modes are complementary in that the first probes parity odd
couplings and the second parity even.  The E871 result on \klmue~
actually predates the time frame of this review, but it is the best
limit available:
BR(\klmue) $< 4.7\times 10^{-12}$\cite{klmue}.  Limits on new
physics are model dependent; it is
typical to derive a limit in a ``generic'' sense for a heavy particle
exchanged at tree level.  For the same coupling strength as the
electroweak couplings of the quarks, the limit on \klmue~ probes
energy scales as high as 150 TeV.\cite{Ritchie}
E865 has completed analysis of its 1998 data set for \kpimue.\cite{Sher}
The dominant background
comes from the overlap of multiple $K^+$ decays, which are estimated
from the time sidebands and extrapolated into the signal region from
the region of high $K^+$ momentum.  Eight events are observed in the
signal region, consistent with background expectations; a likelihood
analysis is used to obtain the branching ratio limit.
Combining this with the E865 results from the
1995 and 1996 runs as well as with the result from the predecessor
experiment BNL E777, yields the final E865 limit: BR(\kpimue) $< 1.2 \times
10^{-11}$.  New results from KTeV are also available on \klpimue
\cite{Bellavance} and \kleemmlfv. \cite{eemumu}

\section{Quark mixing and CP violation: \pnn}

The decays \kpnn~ and \klpnn~ have attracted much attention for their
potential (together with the decay \kethree)
to completely determine the Unitarity Triangle from kaon decays alone.
An inconsistency between the unitarity relation in kaon decays
($s\rightarrow d$ transitions) with that in B decays ($b \rightarrow
d$ transitions) would provide clues to the flavor structure
of physics beyond the SM.
The clean
theoretical nature of the \pnn~ decay modes was discussed at this
conference by Sehgal; a detailed review can be found in
\cite{Buras}. 

First results from BNL E949 on the decay \kpnn~ have been published
recently.\cite{E949}
The signal region is analyzed with a signal-to-noise
likelihood ratio technique.  
An event is seen in
the 2002 dataset near the \kpnn~ kinematic endpoint, albeit with
poorer signal-to-noise ratio compared to the previous two candidate
events seen by E787; accordingly
the new event has an effective contribution to
the branching ratio of about 0.5 events.  The best estimate of the
branching ratio, combining data from E787 and E949 is 
\br(\kpnn) $ = 1.47^{+1.30}_{-0.89}\times 10^{-10}$,
consistent with the SM, although the central value remains
about twice the SM value.
Further details can be found in the presentation by Sekiguchi at this
conference.
A result from E787 on \kpnn~ from the 1997 dataset in
the $\pi^+$ momentum region below 195 MeV/c has also been published
recently.\cite{pnn2_97}

A model-independent bound on the branching ratio for \klpnn~ can be
obtained from the \kpnn~ branching ratio.\cite{Grossman-Nir}  Using
the most recent result from E949, this so-called Grossman-Nir bound
becomes \br(\klpnn) $< 1.4 \times 10^{-9}$, about 400 times
lower than the best direct limit from KTeV.\cite{KTeV97}. As discussed
at this conference by Komatsubara, the first
experiment (KEK E391a) dedicated to studying this decay mode took its
first data this year, hoping to cover the entire region between the KTeV and
Grossman-Nir limits.

\section{Quark mixing and CP violation: other decay modes}

The decays \klpill~ (where $l = e,\mu$) have also attracted interest
for their potential to determine the Wolfenstein
parameter $\eta$, responsible for CPV in the SM.
However, while \klpnn~ proceeds almost entirely due to ``direct'' CPV
(sensitive to $\eta$), the \klpill~ modes have large contributions to
the branching ratio from ``indirect'' CPV ($K_L-K_S$ mixing, followed
by CP conserving (CPC) $K_S$ decay) and from
the interference between direct and
indirect CPV. In the case of \klpimm, there is also a large
contribution from the CP conserving amplitude.
Much theoretical
effort has gone
into disentangling the various contributions as was covered briefly by
Smith, and discussed in detail in \cite{klpill1,klpill2}.

On the experimental side, KTeV has updated the search for
\klpiee, adding the results of their 1999 dataset.\cite{KTeV_piee}
One event is seen, consistent with expectations from background,
dominated by \kleegg.
Combining with the previous result from the 1997 dataset, yields the
final KTeV limit: \br(\klpiee) $< 2.8 \times 10^{-10}$.  This is
still about a factor of 10 above the SM prediction; to beat down the
\kleegg~ background further requires higher precision
tracking and calorimetry, which may be difficult considering that KTeV
is already state-of-the-art. An interference
analysis might be a possible way out.\cite{klpill1}
\klpimm~ has
a less severe background problem from \klmmgg~, but the SM
branching ratio is smaller by about a factor of two.  Results from the
1999 dataset of KTeV are awaited; this data sample contains about
a factor 1.3 more K decays compared to the 1997 data.

The measurements of \kspill~ are important
inputs to the computation of
the contributions of indirect CPV (and hence also the magnitude of the
interference term, but not the sign)
to \klpill.  As shown by Ruggiero, NA48/1 has made the
first observation of both
\kspiee~ and \kspimm:
\br(\kspiee)$= (5.8^{+2.8}_{-2.3}(stat) \pm 0.8 (sys) )
\times 10^{-9}$\cite{kspiee} and
\br(\kspimm)$= (2.9^{+1.4}_{-1.2} \pm 0.2) \times
10^{-9}$.\cite{kspimm}  The precision of these measurements currently set the
uncertainty in the SM expectation for \klpill~ to around 30\%;
it is thought that the ultimate theoretical precision could be
brought below 10\%.\cite{klpill1,klpill2}

Much effort has also gone into extracting the Wolfenstein parameter
$\rho$ from \klmumu~ decays.  The decay itself is well-measured; the
difficulty is that the
branching ratio is almost saturated by the two-photon intermediate state,
masking the short-distance contribution sensitive to $\rho$.  The
imaginary part of the two-photon amplitude can be obtained from
\br(\klgg), as has been known for many years.\cite{Sehgal69}   The real
part can be constrained by studies of the form factor for
$K_L$ decays to virtual photons, with the final states $e^+e^-\gamma$,
$e^+e^-e^+e^-$, $\mu^+\mu^-\gamma$, and $\mu^+\mu^-e^+e^-$.  Studies
of the $ee\gamma$, $eeee$ \cite{LaDue} and $ee\mu\mu$ \cite{eemumu}
states have been updated by KTeV, now
including their full dataset from the 1997 and 1999 runs.  These KTeV
form factor measurements are consistent with one another, but the more
precise measurements ($ee\gamma$ and $\mu\mu\gamma$) are in
disagreement with the previous NA48 measurement from
$ee\gamma$.\cite{Ceccucci} New results are expected from NA48 on
$ee\gamma$ and $eeee$; a measurement of the $eeee$ branching ratio
was shown at this conference by Ruggiero.  Using the KTeV measurements
as input, limits on $\rho$ from \klmumu~ have been
derived.\cite{Isidori}  While these limits are not competitive with
other CKM constraints, and are not expected to improve significantly
without new theoretical ideas, they do provide constraints on
non-standard scenarios.

KLOE is starting to get into the rare decay regime with a new result
on the CP-violating decay \ksppp, as reported by Martini.
The motivation
for this decay mode is that
the uncertainty on the \ksppp~ amplitude currently limits the
precision on $Im(\delta)$, where $\delta$ parametrizes the
CPT-violating part of the $K_L,K_S$ wavefunctions.\cite{CPT}
The result, $\mathcal{B}$(\ksppp) $<2.1 \times
10^{-7}$, is about a factor 70 improvement over the current PDG limit,
and improves the precision on $Im(\delta)$ by about a factor of 2.5.
NA48/1 probes separately the real and
imaginary parts of the \ksppp~ amplitude by looking at $K_L-K_S$
interference, with a similar sensitivity to $Im(\delta)$.\cite{NA48ksppp}

\section{Outlook and Conclusion}

Concerning LFV, existing data on the most sensitive
modes  are now fully analyzed.  There are no currently running or proposed
LFV experiments in the kaon sector;
current methods have been estimated to give perhaps another factor
of 40 at best.\cite{Molzon}
Attention has turned instead to
the muon sector where sensitivity gains of 3-4 orders of magnitude are
anticipated.\cite{mulfv}  SUSY models
generally put LFV far out of reach of kaon experiments while large
parts of parameter space would be accessible by muon decays.\cite{Belyaev}
On the other hand, LFV K decays can probe interesting areas of
parameter space in ETC models.\cite{Applequist}  More generally, LFV K
decays involve both quarks and leptons and could provide information
complementary to that obtained in the muon sector.

Concerning precision tests of the CKM matrix, future efforts are
concentrated on the ``golden'' modes: \kpnn~ and \klpnn.  The current
experimental situation on \kpnn~ cries out for completion of the BNL E949
program (only 20\% of the proposed running has been completed), but the DOE
has halted HEP operations at the AGS; a proposal to NSF to complete
E949 has been
submitted.  In addition, experiments are under consideration
at other labs to take the
sensitivity one step further to the level of between 50 and 100
events with S/N=10.  These include decay-in-flight experiments P940
at Fermilab,
\cite{P940} NA48/3 at CERN, \cite{NA48-3} and a stopped kaon
experiment at J-PARC.\cite{JPARCpnn}  For \klpnn, KEK E391a may have
another run in 2005, possibly going below the Grossman-Nir limit.
The KOPIO experiment \cite{KOPIO} expects to observe 40 events with S/N=2; the
project was included in the FY05 President's Budget for a construction
start in 2005.  A 5-year construction is envisaged with test runs
starting in 2008.  There is also a letter of intent to use
the E391a technique at J-PARC.\cite{JPARCkl}

In conclusion, rare kaon decays continue to be an active area of
study.  LFV decays have reached single-event sensitivities at the
$(2-5)\times 10^{-12}$ level, but
further progress requires new measurement techniques.
Heroic efforts have been made towards understanding the short distance
components of \klpill~ and \klmumu.  Current experiments for
\klpill~ are within striking distance of the SM by a factor of
about 10, but backgrounds are severe. 
Still, the potential exists for the discovery and study of
BSM effects complementary to
those in \pnn; clever ideas for experiments are needed.
The focus of the community is converging on the \pnn~
decays for precision CKM tests.  There are many ideas for experiments
at various labs; some even appear to be funded.  Together with
precision measurements 
at B factories, these could provide decisive tests of the flavor
sector.  It is worth recalling that at least in the movie, the
``Good'' was able to tease out the secret from the
``Ugly''.\footnote{But keep in mind the twist at the very end!}

\section{Acknowledgements}
Thanks to S. Kettell for comments on the draft.
This work was supported in part under US Department of Energy contract
DE-AC02-98CH10886. 

\end{document}